\documentclass[11pt]{article}
\usepackage{geometry}
\usepackage{amssymb}
\usepackage{fontenc}
\usepackage{times}
\usepackage{mathptmx}
\usepackage{graphicx}

\def\thintablerule{\hrule height0.4pt}

\newcommand{\be}{\begin{equation}}
\newcommand{\ee}{\end{equation}}
\newcommand{\bea}{\begin{eqnarray}}
\newcommand{\eea}{\end{eqnarray}}
\newcommand{\bal}{\begin{aligned}}
\newcommand{\eal}{\end{aligned}}

\newcommand{\eq}[1]{Eq.~(\ref{#1})}
\newcommand{\fig}[1]{Fig.~\ref{#1}}

\begin{document}

\centerline{\LARGE Towards the effective action of Non-Perturbative Gauge-Higgs Unification}
\vskip .5cm
\centerline{\LARGE (or on RG flows near quantum phase transitions)}
\vskip .5cm

\vskip 2 cm
\centerline{\large Nikos Irges}
\vskip 1cm
\centerline{\it Department of Physics}
\centerline{\it National Technical University of Athens}
\centerline{\it Zografou Campus, GR-15780 Athens, Greece}

\centerline{\it e-mail: irges@mail.ntua.gr}

\vskip 2.2 true cm
\thintablerule
\vskip 2.0ex
\centerline{\bf Abstract}

We make a few general comments on the Renormalization Group flows in certain Yang-Mills theories in the vicinity of phase transitions.
We then present a model in $d=5$ with non-periodic boundary conditions where a possible RG flow starts from the trivial point and ends 
on a quantum phase transition. Near the endpoint of the flow interesting comments can be made about the Higgs hierarchy problem.
\vskip 1.0ex\noindent
\vskip 2.0ex
\thintablerule

\newpage

%---------------------------------------------------------------------------------------------------------------------------------------------------
\section{Introduction}
%---------------------------------------------------------------------------------------------------------------------------------------------------

Let us consider pure $SU(N)$ Yang-Mills (YM) theories in $d> 4$ dimensions. Such theories are perturbatively non-renormalizable but lattice regularizable.
If there are $n+1$ dimensionful parameters, the phase diagram is $n$-dimensional, equal to the number of dimensionless ratios or "couplings" that one can form.
In a lattice regularization the number of nodes in a given direction is one of these couplings. For example in the simplest (periodic) case it governs
finite temperature phase transitions. A mass in lattice spacing units is another dimensionless coupling and a dimensionless matter, gauge or matter-gauge coupling is another.

We now make two assumptions. The first is that when a phase transition (PT) is approached along a physical trajectory on the phase diagram the system tends to become scale invariant.
The second is that phase transitions possess "shadows", meaning that when the nature of the system changes 
under the tuning of certain couplings, the new system 
has some memory of the physics that the original presence of the phase transition used to impose on it before the tuning took place.

These properties have far reaching consequences for systems that obey them. One immediate consequence of the first assumption is that the lattice spacing decreases towards a PT,
independently of its order. In particular when it is of second order the continuum limit can be taken and the effective theory in many cases is a Conformal Field Theory (CFT).
When it is of first order the lattice spacing does not go to zero, the effective theory is one with a finite cut-off and therefore not a CFT; 
instead it is a theory with (spontaneously) broken conformal symmetry. 
In order to start appreciating the second postulate let us first consider $SU(2)$ YM in $d=5$ at zero temperature. Regularizing on a 5d isotropic and periodic hypercubic lattice
reveals a first order quantum phase transition separating a Confined from a Coulomb phase. Now let us start shrinking the fifth dimension. On the lattice this
can be done only in discrete steps, by decreasing the number of lattice nodes $N_5$. Nevertheless, below some critical $N_5$, 
the PT disappears and the system starts behaving four-dimensional, but with Kaluza-Klein states.
For $N_5=0$ it becomes exactly 4d by definition, with only a Confined phase. The shadow of the 5d PT then involves a dynamically generated scale, say $\Lambda_{\rm D}$.
The scale breaks the classical conformal symmetry of the 4d theory and separates a strong coupling regime from an asymptotically free regime.
This was an example where the tuning of the coupling governing the system's dimensionality, removes the PT. 

Another, but more elusive consequence of the combination of the two postulates is a severe constraint on the dynamics near a PT.
Let us assume that we are on the renormalized trajectory $AC$ on \fig{PD} somewhere near $C$, a point on the PT. 
Because of our first postulate there is some other point on the trajectory $BC$ on the other side of the PT that has the same value(s)
of the lattice spacing(s). Then if the relations between the physical observables defining the trajectories and the lattice spacings on either side is (functionally) invertible, we will have the relation
\begin{equation}\label{DE}
A = P_1\cdot S_1 = P_2\cdot S_2\, .
\end{equation}
In the above $A$ is the $1\times q$ dimensional matrix of lattice spacings (we take into account the possibility of anisotropic lattices with $q$ different lattice spacings), $P_i$, $i=1,2$ is a $1\times p_i$ dimensional matrix of observables
defining the trajectory on the side of the PT labelled by $i$ and $S_i$ is a $p_i\times q$ dimensional matrix, function of the dimensionless couplings.
\eq{DE} can easily be over-constrained  so it is non-trivial if there is a solution. For physical systems it is however self evident that there should be one.
The conclusion is that the dynamics on the two sides are necessarily correlated. 
This becomes particularly interesting when the phases on the two sides of the PT are very different
and the PT is of second order. Then, the dynamics of the system on either side must be such that on the PT (point $C$ in \fig{PD}) they both
end up being describable by the same CFT. The situation is not less interesting when the PT is of first order. In order to comprehend a bit better the
situation let us assume that phase 1 where point $A$ is located is a Higgs phase, while phase 2 where point B is located is a Confined phase. Then 
the effective theory describing the Confined phase possesses a dynamical scale by dimensional transmutation which constrains the
value of the cut-off in the effective theory describing the Higgs phase near the PT. This does not allow the cut-off in the effective Higgs sector
take too large values, as the maximum value of the cut-off on the PT is constrained by the intrinsically low value of the $\Lambda_{\rm D}$ parameter of the Confined phase.
This fact, together with the rather natural tuning of the couplings that is necessary to perform so that one remains on the physical trajectory can be viewed 
as an alternative (e.g. to supersymmetry) resolution to the Higgs sector's naturalness problem.

To appreciate even more the constrained dynamics near phase transitions that follows from our two postulates, let us try to stretch the situation to its extreme.
Let us imagine that instead of a maximum number of three phases that meet at the "triple" point $M$ on \fig{PD} there are many more phases meeting at some point
on the phase diagram and we can approach this multiple meeting point along physical trajectories from any phase. Then the system when brought near the multiple point should "feel" the
presence of the other nearby phases and its dynamics is more constrained the more phases meet at the multiple point.
This translates into constraints on the quantum behaviour of its physical observables via the first of our postulates. 
Note that the constrained dynamics is achieved in this case by dynamics as opposed to constrained dynamics obtained by increasing the global symmetries.
This is just the multiple point principle of \cite{MultiplePoint}, expressed here as a consequence of our first postulate.

Now global symmetries also constrain the dynamics of a system. One of the many ways to increase the global symmetries is to add extra matter
with appropriate couplings in a way that makes the system supersymmetric or to tune couplings so that the operation changes its dimensionality. 
Such operations however (we saw an example above) may alter the phase diagram itself at the same time
and it could happen that some or all of the phase transitions may disappear. 
The shadow of the disappearing phase transitions in this case may include some sort of a strong-weak coupling duality.
Moreover if at the end all that remains on the phase diagram is an isolated point of second order phase transition, we may have a
supersymmetric CFT with very constrained dynamics.

Next we present a model where at least some of the basic features presented above can be demonstrated.

%---------------------------------------------------------------------------------------------------------------------------------------------------
\section{NPGHU and its effective action}
%---------------------------------------------------------------------------------------------------------------------------------------------------

\begin{figure*}[!ht]
\centerline{\includegraphics[width=100mm]{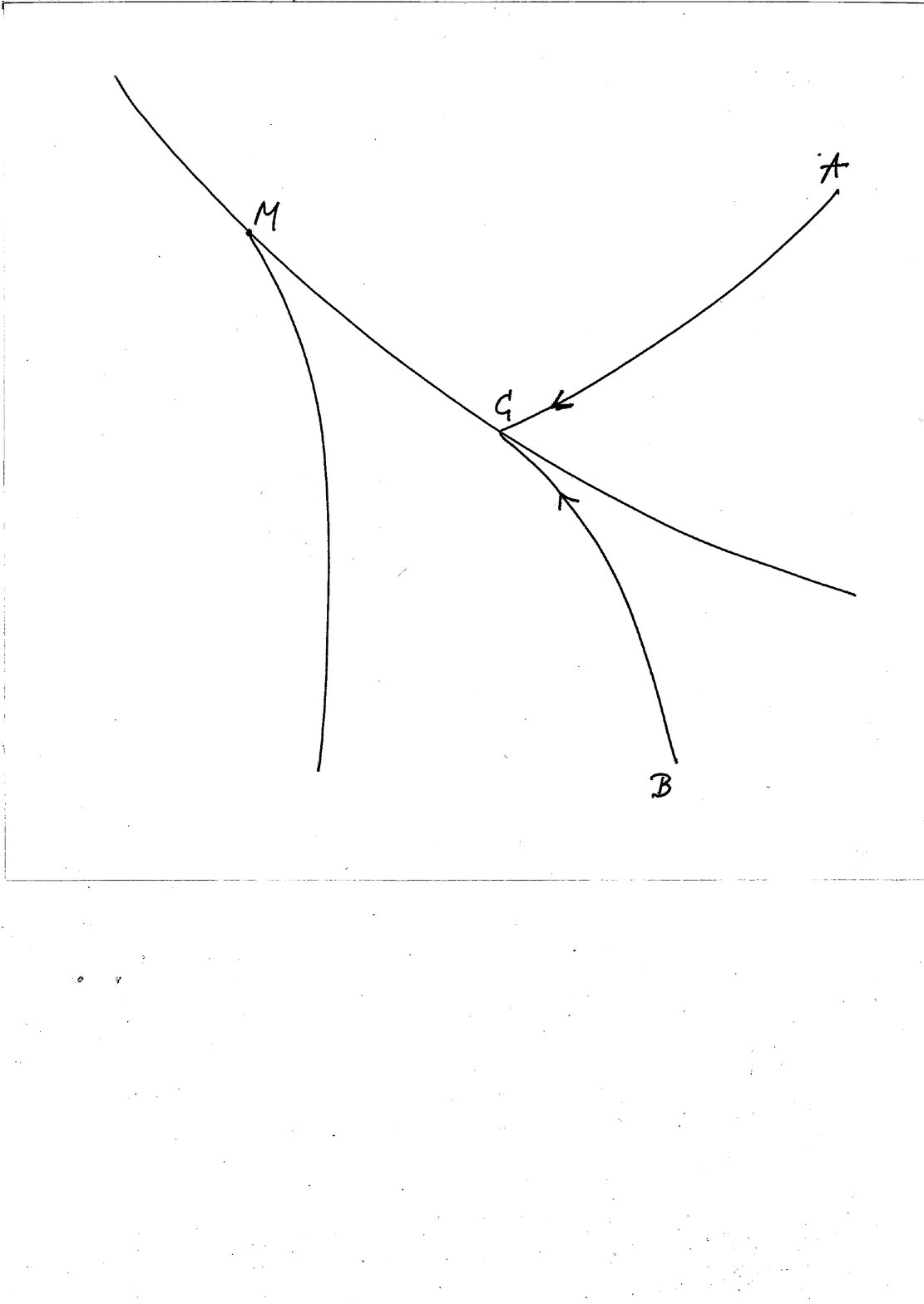}}
\caption{\small{Schematic phase diagram of the $5d$ orbifold lattice model.
\vspace{0cm}}}\label{PD}
\end{figure*}

In \cite{Corfu16} we anticipated some of the properties described in the previous section using a particular lattice model \cite{Orblat}.
We call this model one of "Non-Perturbative Gauge-Higgs Unification (NPGHU)" because the Brout-Englert-Higgs (BEH) mechanism involved is a quantum effect and can be 
seen non-perturbatively on the lattice. The basic construction is a 5d lattice with large and periodic four dimensions each represented by $L$ lattice nodes and a fifth dimension which is an interval. 
In the bulk there is a pure $SU(2)$ gauge theory and at the two ends of the interval, on the 4d boundaries, only a $U(1)$ subgroup coupled to a complex scalar survive.
The dimensionless couplings that parametrize the phase diagram are the number of nodes in the fifth dimension $N_5$, the lattice gauge coupling $\beta$ and the anisotropy $\gamma$.
The anisotropy appears because we have taken the lattice spacing in the four-dimensional sense $a_4$ to be different than the lattice spacing in the fifth dimension $a_5$.
Then $\gamma=a_4/a_5$. We will be interested in the $\gamma < 1$ regime where the phase diagram looks like in \fig{PD} and is rather insensitive to $N_5$.
The phase where point $A$ belongs is a Higgs phase and the one where point $B$ belongs is the "Hybrid" phase. The curves passing through $M$ are lines of phase transitions.
The curves $AC$ and $BC$ are physical trajectories or "Lines of Constant Physics (LCP)". 
Indeed, as the point $C$ is approached from point $A$, all masses decrease in units of $a_4$, confirming the first of our assumptions.
We will be eventually interested to understand the vicinity of the point $C$ where the quantum Higgs-Hybrid phase transition takes place.
In the Higgs phase and near the PT into the Hybrid phase the boundary decouples from the bulk independently of $N_5$, which remains five-dimensional.\cite{Maurizio}
We will call this regime of Dimensional Reduction via Localization (as opposed to compactification) the DRL regime.
In the Hybrid phase the system decomposes into 4d slices which means that on the boundary we have at the classical level a massless scalar QED while in the bulk
an array of 4d $SU(2)$ gauge theories. 
The line of PT in the orbifold phase diagram separating the Confined and Hybrid phases from the Higgs phase (the line that passes through points $M$ and $C$)
is inherited from the periodic 5d system and it is there because of the orbifold's bulk. Upon introducing the non-periodic boundary conditions, the PT line
separating the Confined and Hybrid phases appears. It must be there because in the DRL regime the phase diagram must reproduce the phase diagram 
of the Abelian-Higgs model (with a charge 2 Higgs). As a result, the Abelian-Higgs model inherits the structure of its phase diagram from the 5d orbifold
due to dimensional reduction in the DRL regime. 
To use our terminology, the 4d Abelian-Higgs phase diagram is the shadow of the 5d orbifold phase diagram. 
Note that in order that this is realized it is necessary to introduce an anisotropy parameter in the orbifold lattice.

To proceed with some technical details we have to specify the methods involved in the analysis.
There are two ways to attack the model quantitatively on the lattice. One is via a Mean-Field (MF) expansion and the other via Monte Carlo (MC) simulations.
The former approach should be considered as an approximation to the latter. The advantage of the MF is that it gives a semi-analytical control
and its disadvantage that it sees the Higgs-Hybrid PT as second order while in reality it is of first order, as the MC method reveals.
For all other phase transitions the two methods agree. This gives us the opportunity to turn the failure of the MF prediction to our favour
by considering the effective action near such quantum phase transitions to be a spontaneously broken CFT instead of a real CFT;
the latter is the limit that the leading order MF expansions sees.
It is important to know the dynamically generated Higgs potential on the boundary as the PT is approached.
Within the MF approach this can be made possible by defining observables called "cumulants". These are gauge invariant scalar operators 
built from powers of Higgs-like lattice operators extending in the fifth dimension but starting and ending on the boundary. As such they can be considered as boundary operators.
They are $L$-dependent.
The work of L\"uscher, Weisz and Wolff on "step scaling" \cite{Step} teaches us how $\beta$ functions associated with such finite size couplings can be 
extracted. The method can be actually applied to the model and it will hopefully allow us to "see"
the shape of the effective scalar potential, at least in the vicinity of the PT.\cite{Andreas}

We would like to have also a continuum effective description of the physics near quantum phase transitions.
In this respect, our available tools are very limited at present.
Understanding the physics near the Higgs-Hybrid phase transition
involves at least a partial understanding of the RG flows in the system. Starting from the 5d weak coupling or "trivial fixed point" regime, in an infinite 5d volume,
we have a bulk which is a pure $SU(2)$ gauge theory and a massless scalar QED with vanishing quartic interaction on the boundary.
To leading order in a cut-off expansion the bulk is decoupled from the boundary and at the classical level there is no SSB. 
Regarding the bulk, even though it involves a perturbatively non-renormalizable theory, 
the running of the gauge coupling can be computed.\cite{Dienes}
On the boundary on the other hand, at the classical level we have two separate CFT's, as both a free abelian gauge boson and a free scalar are
CFT's in $d=4$ but the coupled system is not a CFT because non-trivial $\beta$-functions develop at 1-loop. This means that on the boundary, 
in the approximation where it is decoupled from the bulk, we have spontaneous breaking of the conformal symmetry at the quantum level.
At next to leading order, when the coupling of the bulk to the boundary is turned on, we expect in addition the spontaneous breaking of 
the boundary $U(1)$ gauge symmetry. This can be made quantitatively precise only when the bulk to boundary coupling is specified. It should be obvious from 
the previous discussion that we can define this coupling via the allowed by gauge invariance higher dimensional operators of the 5d orbifold
action.\cite{Francesco1} This will basically yield the Wilson operators already mentioned above with coefficients
that can be defined order by order in perturbation theory. Quantizing the
action augmented with the Wilson operators guarantees the breaking of the gauge symmetry, provided that these operators break the global symmetry
that replaces the center symmetry along the orbifold direction.\cite{Stick} 
As a result, we will obtain a continuum effective description of the system near its trivial fixed point that, among other things,
will serve as perhaps the simplest example of a model with a correlated radiative breaking of
conformal and gauge symmetries.  This is a computation currently in progress \cite{Fotis2}, using the $\epsilon$-expansion and exploiting an appropriate limit of the 
gauge invariant 1-loop effective action of the Abelian-Higgs model.\cite{Fotis1}
In the presence of a boundary-bulk coupling SSB will be of course communicated to the bulk.
At the end of the 1-loop computation we will have RG flow lines starting from the vanishing 5d coupling point and extending somewhere
in the interior of the phase diagram, up to the point where the perturbative expansion breaks down. 
Because of perturbative non-renormalizability it is not clear where this point is. In fact this RG flow by itself is rather incomplete.
To make it more meaningful, more has to be done.

One can attempt to do the same but this time starting from the bulk phase transition, in the DRL regime. 
Now in a continuum approach we have an array of 4d slices. The MF and MC analyses instruct us to define an array of weakly interacting softly broken 4d CFT's.
Taking into account that the MF is an approximation to the MC approach,
we can attempt to find first what the array of the 4d CFT's are. Then the broken version with a finite cut-off could arise as it
did near the trivial point, namely via a soft (radiative) breaking parametrized by couplings between the 4d slices. The effect of these couplings should be such that at the end
in the bulk a 5d $SU(2)$ Yang-Mills theory should be recovered while the boundary should remain four-dimensional.
It is amusing to notice that the resulting effective action will bare a similarity to "clockwork" models.\cite{Clockwork}
The RG flow line emanating from the phase transition can now be attempted to be connected to the one emanating from the trivial point.
Global symmetries could be a useful guide for this operation.

%%%%%%%%%%%
%%%%%%%%%%%

\newpage
{\bf Acknowledgements}

We would like to thank the organizers of the 2017 Corfu Summer School and Workshop for the invitation
and A. Chatziagapiou, F. Knechtli and F. Koutroulis for discussions.

\end{document}